\documentclass{article}
\usepackage[utf8]{inputenc}

\usepackage{amsmath} 
\usepackage{amsfonts}
\usepackage{dsfont}
\usepackage[utf8]{inputenc}
\usepackage[english]{babel}
\usepackage[T1]{fontenc}
\usepackage{physics}
\usepackage{tikz}
\usepackage{authblk}

\usepackage{graphicx}
\usepackage{subcaption}
\makeatletter
\newsavebox{\@brx}
\newcommand{\llangle}[1][]{\savebox{\@brx}{\(\m@th{#1\langle}\)}%
  \mathopen{\copy\@brx\kern-0.5\wd\@brx\usebox{\@brx}}}
\newcommand{\rrangle}[1][]{\savebox{\@brx}{\(\m@th{#1\rangle}\)}%
  \mathclose{\copy\@brx\kern-0.5\wd\@brx\usebox{\@brx}}}
\makeatother

\begin{document}

\title{Entanglement of Free Fermions on Hamming Graphs}
\author[1]{Pierre-Antoine Bernard}
\author[2]{Nicolas Crampé}
\author[1]{Luc Vinet}
\affil[1]{Centre de recherches mathématiques, Université de Montréal, P.O. Box 6128, Centre-ville
Station, Montréal (Québec), H3C 3J7, Canada,}
\affil[2]{Institut Denis-Poisson CNRS/UMR 7013 - Université de Tours - Université d’Orléans, Parc de
Grandmont, 37200 Tours, France.}

\maketitle
\begin{abstract}
    Free fermions on Hamming graphs $H(d,q)$ are considered and the entanglement entropy for two types of subsystems is computed. For subsets of vertices that form Hamming subgraphs, an analytical expression is obtained. For subsets corresponding to a neighborhood, i.e. to a set of sites at a fixed distance from a reference vertex, a decomposition in irreducible submodules of the Terwilliger algebra of $H(d,q)$ also yields a closed formula for the entanglement entropy. Finally, for subsystems made out of multiple neighborhoods, it is shown how to construct a block-tridiagonal operator which commutes with the entanglement Hamiltonian. It is identified as a BC-Gaudin magnet Hamiltonian in a magnetic field and is diagonalized by the modified algebraic Bethe ansatz.
\end{abstract}
\section{Introduction}

Entanglement entropy describes the extent to which a subsystem is correlated with its complementary part in a given state of the system. It is a key quantity in quantum information and in the investigation of quantum many-body models \cite{Latorre_2009,Peschel_2012}. Owning to their simplicity, free fermions offer instructive testbeds for its study.

We here examine entanglement in a system of free fermions hopping on the vertices of Hamming graphs $H(d,q)$. This family of graphs has been extensively analyzed and includes the complete graphs $H(1,q)$ as well as the $d-$cubes $H(d,2)$. 

In algebraic combinatorics, Hamming graphs provide the archetype of a $P-$ and $Q-$ association scheme and play a crucial role in coding theory \cite{Bannai1984AlgebraicCI, brouwer2012distance}. The algebra spanned by the adjacency and dual adjacency matrices of $H(d,q)$ is referred to as its Terwilliger algebra $\mathcal{T}$ \cite{GO2002399, LEVSTEIN20061, thini,thinii,thiniii}. Most objects that we shall use to determine the entanglement entropy arise from this structure.

Hamming graphs are known to admit both perfect state transfer \cite{Christandl_2005} and fractional revival \cite{Bernard_2018}. The entanglement entropy of coupled harmonic oscillators on Hamming networks was considered in \cite{jafarizadeh2014entanglement, Jafarizadeh2015EntanglementEI}, where it was shown that the Schmidt numbers and thus the entropy could be extracted from the potential matrix of the system expressed in the right basis. Looking at free fermions, we shall use the same basis, which is tied to the irreducible representations of the Terwilliger algebra of the Hamming scheme. 

We shall examine the entanglement entropy when the system is in its ground state. For free fermions, it is known that the spectrum of the chopped correlation matrix $C$ suffices to compute the entanglement entropy \cite{Peschel_2003, Peschel_2009}. We shall look for the eigenvalues of $C$ for two types of subsystems.

First, we consider subsets of vertices that form by themselves a Hamming graph. For instance, it may correspond to a $2-$cube in a $3-$cube, i.e. to one of the faces of a three dimensional hypercube. For such cases, a general formula for the eigenvalues of $C$, their degeneracy and the entanglement entropy can be derived. 

Second, we examine neighborhoods. A neighborhood is the set of vertices at a given distance from a reference site. For this type of subsystem, the restricted or chopped correlation matrix can be expressed in terms of elements in $\mathcal{T}$, the Terwilliger algebra of $H(d,q)$. The decomposition in irreducible $\mathcal{T}$-submodules of the vector space on which $C$ is acting will reduce our system of free fermions on a graph to a direct sum of fermionic chains (or paths). For a subsystem composed of a single neighborhood, this decomposition diagonalizes $C$ and allows a direct computation of the entanglement entropy. For a subsystem composed of a large number of neighborhoods, it is however not sufficient. Nevertheless, we can use a method recently introduced in \cite{crampe2020entanglement} to contruct a block-tridiagonal operator $T$, referred to as a generalized Heun operator, which commutes with the chopped correlation matrix. This approach is an application to the computation of entanglement entropy of the tools associated to time and band limiting problems  \cite{Cramp__2019,Cramp__2020,Landau1985,Slepian1983}. For the systems considered here, the generalized Heun operator corresponds to a BC-Gaudin magnet Hamiltonian in a magnetic field and it proves possible to diagonalize this operator using the modified algebraic Bethe ansatz \cite{bernard2020heun,Crampe_2017}. This done, the eigenvalues of $C$ and the entropy can be extracted.

The paper is organized as follows. In section \ref{s1} and \ref{s2}, we introduce the Hamiltonian associated to free fermions on a Hamming graph. We diagonalize it, define its ground state and discuss entanglement entropy. In section \ref{s3}, we obtain the entropy for subsystems corresponding to Hamming subgraphs. In section \ref{s4}, we consider subsystems associated to the bundles of neighborhoods. We give an overview of the Terwilliger algebra of the Hamming scheme and of its relevance to the problem at hand. We present the decomposition in irreducible $\mathcal{T}$-submodules and explain the simplification this brings.  $\mathcal{T}$ will also be seen to be the quotient of the Onsager algebra \cite{onsagerDavies,onsager,roan1991onsager} by certain Davies relations \cite{Davies_1990}. We show how the eigenvalues of the chopped correlation matrix $C$ are obtained when the subsystem is a single neighborhood. The generalized Heun operator $T$ is introduced in the subsection \ref{s5}. We indicate that it commutes with $C$ and diagonalize it using the modified algebraic Bethe ansatz. 

\section{Free fermions on Hamming graphs}
\label{s1}

The Hamming graph $H(d,q)$ can be constructed in the following way. Take the set of vertices $F_q^d$ to consist of all the $d$-tuples $v = (v_1, v_2, \dots, v_d)$ for which the elements are integers in $\{0, 1, \dots, q-1\}$. $v$ and $v'$ are connected by an edge when there exists a unique position $i$ in the $d$-tuples such that $v_i \neq v_i'$. It follows that the distance between two vertices in the resulting graph is given by the Hamming distance $\partial$ between their associated tuple:
\begin{align}
    \partial(v,v') = \# \{i \in \{1, \dots,d\}\ |\  v_i \neq v_i'\}.
\end{align}
\noindent It is worth noting that the case $d = 1$ corresponds to the complete graph $K_q$ and that the case $q = 2$ leads to the $d-$cube. We shall consider systems of free fermions living on the vertices of a Hamming graph. We shall study in particular fermionic systems for which the hopping constant between the sites $v$ and $v'$, $\alpha_{\partial(v,v')} \in \mathbb{R}$, depends only on their Hamming distance $\partial(v,v')$. More precisely, the Hamiltonian is defined as
\begin{align}
    \widehat{\mathcal{H}} = \sum_{v, v' \in F^d_q } \alpha_{\partial(v,v')} c_{v}^\dagger c_{v'},
    \label{hamil}
\end{align}
\noindent where $c_v^\dagger$ and $ c_{v}$ are fermionic creation and annihilation operators associated to the site $v$ in the graph. They satisfy the following canonical relations:
\begin{align}
    \{c_v, c_{v'}\} = 0, \quad \{c_v^\dagger,c_{v'}^\dagger\} = 0 \quad \text{and} \quad  \{c_v,c_{v'}^\dagger\} = \delta_{v v'}, \quad \quad\text{for } v,v' \in F^d_q.
    \label{ferm}
\end{align}
\noindent Since the diameter of the graph $H(d,q)$ is $d$, the model defined by \eqref{hamil} contains $d+1$ independent parameters. The constant $\alpha_0$ is related to the presence of an external magnetic field. 

\bigskip

We can give an alternative expression for $\widehat{\mathcal{H}}$. For $i \in \{0, 1, \dots, d\}$, one can define the $i^{\text{th}}$ adjacency matrix $A_i$ of $H(d,q)$ as the matrix whose entry $[A_i]_{vv'}$ is
\begin{align}
[A_i]_{vv'} = 
\left\{
    \begin{array}{ll}
    	1  & \mbox{if } \partial(v,v') = i, \\
    	0 & \mbox{otherwise. }
    \end{array}
\right.
\end{align}
\noindent Furthermore, the vertex associated to the tuple $v = (v_1, \dots, v_d)$ can be represented by the following vector in $(\mathbb{C}^{q})^{\otimes d}$:
\begin{align}
    \ket{v} = \ket{v_1} \otimes \ket{v_2} \otimes \dots \otimes \ket{v_d},
\end{align}
\noindent where $\ket{v_i}$ is the vector in $\mathbb{C}^{q}$ which has its $(v_i + 1)^{\text{th}}$ entry equals to $1$ as its unique non-zero entry:
\begin{align}
    \ket{v_i} = (\underbrace{0, 0, \dots, 0}_{v_i \text{ times}}, 1, 0, \dots,0 )^T.
\end{align}
In this basis, the first adjacency matrix is expressed as 
\begin{align}
    A_1 \equiv A = \sum_{i = 1}^d \underbrace{\mathds{1}_{q \times q} \otimes \dots \otimes \mathds{1}_{q \times q}}_{i-1 \text{ times}} \otimes \ (J_{q \times q} - \mathds{1}_{q \times q})\otimes \underbrace{\mathds{1}_{q \times q} \otimes \dots \otimes \mathds{1}_{q \times q}}_{d-i \text{ times}},
    \label{fia}
\end{align}
\noindent where $\mathds{1}_{q \times q}$ is the identity matrix and $J_{q \times q}$ is the matrix of ones. More generally, we have 
\begin{align}
    A_i = \sum_{\substack{b \in F_2^d, \\ \text{wt}(b) = i}} \underbrace{(J_{q \times q} - \mathds{1}_{q \times q})^{b_1}\otimes(J_{q \times q} - \mathds{1}_{q \times q})^{b_2} \otimes \dots \otimes (J_{q \times q} - \mathds{1}_{q \times q})^{b_{d}}}_{d \text{ times}},
\end{align}
 \noindent where the weight of a binary string $b$ is $\text{wt}(b) \equiv \#\{n : b_n \neq 0 \} = \sum_{n=1}^{d}b_n$. Using the vectors of operators $\hat{c}^\dagger = \sum_{v \in F_q^d} c^\dagger_v\bra{v}$ and $\hat{c} = \sum_{v \in F_q^d} c_v \ket{v }$, the Hamiltonian can be rewritten as
\begin{align}
    \widehat{\mathcal{H}} = \hat{c}^\dagger\Big[ \sum_{i = 0}^d \alpha_i A_i \Big] \hat{c}\ .
    \label{hamil2}
\end{align}

\subsection{Diagonalization and energies}

\noindent We are now interested in obtaining the single-particle excitation energies of the system. Diagonalizing $\widehat{\mathcal{H}}$ amounts to diagonalizing $\sum_{i = 0}^d \alpha_i A_i$, and in fact $A_1$. Indeed, since the Hamming graphs are distance-regular, it is known from the theory of association schemes that the matrix $A_i$ is a polynomial of degree $i$ of the first adjacency matrix $A$ \cite{Bannai1984AlgebraicCI}. Specifically,
\begin{equation}
    \begin{split}
    A_i &= \binom{d}{i} (q-1)^i\sum_{j=0}^i \frac{\left(-i\right)_j \left(-\frac{(q-1)d}{q}+ \frac{A}{q}\right)_j}{\left(-d\right)_j\  j!} \Big(\frac{q}{q-1}\Big)^j\\
    &= \binom{d}{i} (q-1)^i K_i\left(\frac{(q-1)d}{q}- \frac{A}{q} ;\frac{q-1}{q},d \right),
    \label{aia1}
    \end{split}
\end{equation}
\noindent where $(a)_i = (a)(a+1)\dots(a+i-1)$ and $K_i$ refers to the Krawtchouk polynomial of degree $i$ \cite{koekoek1996askey}.  Therefore, diagonalizing all the $d+1$ adjacency matrices is the same as diagonalizing $A$. This can be achieved by considering tensor products of $d$ eigenvectors of the matrix $J_{q \times q}$. This matrix has a unique eigenvector $\ket{\theta_q}$ of eigenvalue $q$:
\begin{align}
    \ket{\theta_q} = \frac{1}{\sqrt{q}}\sum_{i = 0}^{q-1}\ket{i},
\end{align}
and its other eigenspaces have eigenvalue $0$ and a degeneracy of $q-1$. We take $\{\ket{\theta_i} : i \in \{1,\dots,q\}\}$ to be an orthonormal basis of this space. Thus, we have $\bra{\theta_i}\ket{\theta_j} = \delta_{ij}$,
\begin{align}
    J_{q \times q} \ket{\theta_q} = q \ket{\theta_q} \quad \text{and} \quad J_{q \times q} \ket{\theta_i} = 0 \quad \text{if } i \neq q. 
\end{align}
\noindent Then, if $\ket{\theta_{i_1} \theta_{i_2} \dots \theta_{i_d}} = \ket{\theta_{i_1}} \otimes \ket{\theta_{i_2}} \otimes \dots \otimes \ket{\theta_{i_d}}$ with $i_1, \dots, i_{d-1}$ and $i_d$ $\in \{1,\dots, q\}$, we see from \eqref{fia} that
\begin{align}
    A \ket{\theta_{i_1} \theta_{i_2} \dots \theta_{i_d}} = (k q - d) \ket{\theta_{i_1} \theta_{i_2} \dots \theta_{i_d}},
    \label{eg}
\end{align}
\noindent where $k$ is the number of $\ket{\theta_q}$ in the tensor product. The spectrum of $A$ is thus $\omega_k = kq - d$ with $k \in \{0, 1, \dots, d\}$. As for the degeneracy of the $k^{th}$ eigenspace, it is easy to see that it is given by $D_k = \binom{d}{k}(q-1)^{d-k}$. Let us also note that since $\ket{\theta_q}\bra{\theta_q} = J_{q \times q}/q$, the projection operator $E_{k}$ over the $(d-k)^{th}$ eigenspace of $A$ can be expressed as
\begin{align}
    E_{k} = \sum_{\substack{b \in F_2^d, \\ \text{wt}(b) = k}}  {(b_1 \mathds{1} + (-1)^{b_1}\frac{J_{q \times q}}{q})\otimes\dots \otimes(b_{d}\mathds{1} + (-1)^{b_{d}} \frac{J_{q \times q}}{q})}.
    \label{projk}
\end{align}
From the spectrum of $A$ and \eqref{aia1}, we can deduce that the eigenvalues of $\sum_{i = 0}^d \alpha_i A_i$, i.e. the single-particle excitation energies $\Omega_k$ of the system, are 
\begin{align}
    \Omega_k = \sum_{i=0}^d \alpha_i \binom{d}{i} (q-1)^i K_i( d-k ;\frac{q-1}{q},d ).
    \label{ener}
\end{align}
\noindent Indeed, if we take the set $\{\ket{\omega_k,l}\}$ ($l$ labels the degeneracy) to be an orthonormal basis of eigenvectors of $A$, we have
\begin{align}
    \widehat{\mathcal{H}} = \sum_{k = 0}^d\sum_{l = 1}^{D_k} \Omega_k \Bar{c}_{kl}^\dagger \Bar{c}_{kl} ,
\end{align}
\noindent where $\Bar{c}_{kl}^\dagger  = \sum_v \bra{v}\ket{\omega_k,l} c_v^\dagger $ and $\Bar{c}_{kl}  = \sum_v \bra{\omega_k,l}\ket{v} c_v $. One can check that these new creation and annihilation operators respect the same canonical relations as the operators $c_v$ and $c_{v'}^\dagger$. Formula \eqref{ener} reduces to a simpler form in some useful cases. For instance, if we restrict ourselves to the nearest neighbour interactions ($\alpha_i = 0$ for $i \neq \{0,1\}$), the energies are linear in $k$:
\begin{align}
    \Omega_k = \alpha_0 + \alpha_1 (kq - d) \ .
\end{align}
\noindent Furthermore, in the case where the hopping constants decrease exponentially with the distance, we have $\alpha_i = e^{-ci}$ for $i > 0$, where $c \geq 0$. Thus, we find that \eqref{ener} corresponds to the generating function of the Krawtchouk polynomials which grows exponentially with $k$ :
\begin{align}
    \Omega_k &= (1  - e^{-c})^{d-k} (1+ e^{-c}(q-1))^{k} + \alpha_0 - 1 .
\end{align}

\subsection{Ground state}

\noindent Let $|0 \rrangle$ be the vacuum state annihilated by all the operators $\bar{c}_{kl}$. The ground state $|\Psi_0 \rrangle$ is the state for which all the energy levels $\Omega_k < 0$ are occupied, which corresponds to filling up the Fermi sea. We denote $SE$ the set of all the integers $k \in \{0, 1, \dots, d\}$ associated to a negative single-particle excitation energy $\Omega_k$. For some fixed parameters $\alpha_i$, one can easily identify $SE$ by computing the values taken by \eqref{ener}. In the case where the hopping terms decrease exponentially with the distance, we know from the last section that it corresponds to a set of the form $\{0,1, \dots, k_0\}$ for some integer $k_0$. In any case, we have:
\begin{align}
    |\Psi_0 \rrangle = \Big[\prod_{k\in SE} \ \prod_{l = 1}^{D_k}\bar{c}_{kl}^\dagger \Big]  |0 \rrangle .
\end{align}
\noindent As we will see in the next section, the key information we need is contained in the correlation matrix $\widehat{C}$ whose components $\widehat{C}_{v v'}$ are defined as 
\begin{align}
    \widehat{C}_{v v'} = \llangle \Psi_0 | c_v^\dagger c_{v'}|\Psi_0 \rrangle, \quad \quad \text{where }  v, v' \in F^d_q.
\end{align}
\noindent We can use the eigenbasis of $A$ to express $c_v^\dagger$ and $c_v$ in terms of $\Bar{c}_{kl}^\dagger$ and $\Bar{c}_{kl}$. Then, simple algebraic manipulations are sufficient to show that 
\begin{equation}
\begin{split}
     \widehat{C} &= \sum_{k \in SE} \sum_{l = 1}^{D_k} \ket{\omega_k,l}\bra{\omega_k,l} \\
    &= \sum_{k \in SE} E_{d-k} \equiv \pi_{SE},
    \label{proen}
\end{split}
\end{equation}
\noindent where $E_{d-k}$ is the projection operator onto the $k^{th}$ eigenspace of $A$ given by \eqref{projk} and $\pi_{SE}$ is the projection operator onto all the eigenspaces associated to an integer in $SE$. 

\section{Entanglement entropy}
\label{s2}
Let $SV \subset F^d_q$ be a subset of sites of the Hamming graph referred to as the subsystem $1$. The projection operator over this subsystem is 
\begin{align}
    \pi_{SV} = \sum_{v \in SV} \ket{v}\bra{v}.
    \label{propoz}
\end{align} 
We shall refer to its complement $F^d_q \backslash SV$ as the subsystem $2$. In the ground state, the reduced density matrix of the subsystem $1$ is defined by
\begin{align}
    \rho_1 = \text{tr}_2 |\Psi_0 \rrangle \llangle \Psi_0 |
\end{align}
\noindent and its von Neumann entropy $S$ is
\begin{align}
    S = - \text{tr}(\rho_1 \ln{\rho_1}).
\end{align}
\noindent This quantity allows one to determine to which degree the state of $SV$ is interwined with the rest of the system. Once the eigenvalues of $\rho_1$ are determined, its computation is easy to realize. It is known that these eigenvalues can be extracted from those of the chopped correlation matrix $C$ \cite{Peschel_2003,Peschel_2009}, which is defined as
\begin{align}
    C = |\widehat{C}_{v v'}|_{vv' \in SV}
\end{align}
\noindent and given by 
\begin{align}
    C = \pi_{SV}  \pi_{SE}  \pi_{SV},
\end{align}
\noindent in terms of the projectors \eqref{proen} and \eqref{propoz}. In a nutshell, since the ground state is a Slater determinant, all the correlations can be expressed in terms of two-particle functions. This imply by Wick's theorem that the reduced density matrix $\rho_1$ is expressible as the exponential of a quadratic operator referred to as the entanglement hamiltonian $\mathcal{H}$:
\begin{align}
    \rho_1 = \kappa \exp(-\mathcal{H}), \quad \quad \text{where} \quad \quad \mathcal{H} = \sum_{v,v' \in SV} h_{vv'} c_v^\dagger c_{v'}.
\end{align}
\noindent Since $C_{v v'} = \text{tr}(\rho_1\ c_v^\dagger c_{v'} )$, we see that $C$ and $h$ can be simultaneously diagonalized and that their spectra are related by the relation
\begin{align}
    h = \ln{[(1-C)/C]}.
\end{align}
\noindent Following \cite{Carrasco_2017}, we can give the entropy in terms of the eigenvalues $\lambda$ of $C$ and their degeneracy $D_\lambda$:
\begin{align}
    S = - \sum_{\lambda} D_\lambda \left[\lambda \ln{(\lambda)} + (1-\lambda)\ln{(1-\lambda)}\right].
    \label{entropy}
\end{align}
\noindent In the following, we aim to diagonalize $C$ in two interesting cases.

\section{Entanglement entropy for Hamming subgraphs}
\label{s3}
\begin{figure}[h]
    \centering
    \begin{tikzpicture}
\draw[line width=0.25mm, black ] (-3,0) -- (-1,1.5);
\draw (-3,0) -- (-1,0);
\draw[line width=0.25mm, black ] (-3,0) -- (-1,-1.5);

\draw (3,0) -- (1,1.5);
\draw (3,0) -- (1,0);
\draw (3,0) -- (1,-1.5);

\draw (-1,1.5) -- (1,1.5);
\draw[line width=0.25mm, black ] (-1,1.5) -- (1,0);

\draw[line width=0.25mm, black ] (-1,-1.5) -- (1,0);
\draw (-1,-1.5) -- (1,-1.5);

\draw (-1,0) -- (1,1.5);
\draw (-1,0) -- (1,-1.5);

\draw[black,fill=white] (1,1.5) circle (0.2cm);
\node[] at (1.7,1.7) {110};
\draw[black,fill=black] (1,0) circle (0.2cm);
\node[] at (1.7,0.2) {011};
\draw[black,fill=white] (1,-1.5) circle (0.2cm);
\node[] at (1.7,-1.5) {101};

\draw[black,fill=black] (-1,1.5) circle (0.2cm);
\node[] at (-1.7,1.7) {010};
\draw[black,fill=white] (-1,0) circle (0.2cm);
\node[] at (-1.7,0.2) {100};
\draw[black,fill=black] (-1,-1.5) circle (0.2cm);
\node[] at (-1.7,-1.5) {001};

\draw[black,fill=white] (3,0) circle (0.2cm);
\node[] at (3.7,0.2) {111};
\draw[black,fill=black] (-3,0) circle (0.2cm);
\node[] at (-3.7,0.2) {000};
\end{tikzpicture}

    \caption{A three dimensional cube and one of its Hamming subgraph (in black) that corresponds to $SV_2$, a two dimensional cube.}
    \label{fig:subham}
\end{figure}
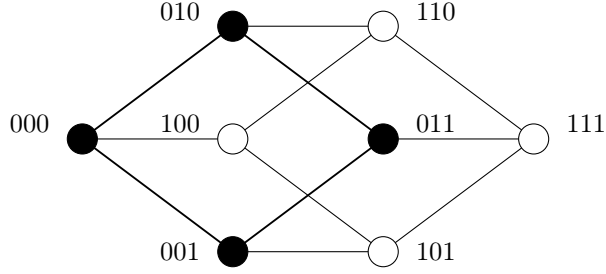

In this section, we take as subsystem $1$ a subset of vertices which forms a Hamming graph of lower dimension. Take the full system to be the Hamming graph $H(d,q)$ associated to tuples of length $d$ constructed from elements in $\{0,1,\dots, q-1\}$. A subset which corresponds to a Hamming subgraph is obtained by fixing some elements in the $d$-tuples.
\noindent Here, without loss of generality, we fix the $d-L$ first elements to zero. The subset of $d$-tuples $SV_L$ corresponding to the subsystem 1 is thus
\begin{align}
    SV_L = \{v \in F^d_q : \ v_i = 0 \ \forall \ i \leq d- L \}.
\end{align}
 For instance, in the case $d =3 $ and $q  = 2$, we could consider the subset of binary tuples starting with one $0$. This is the same as considering one of the faces of a 3-cube (see Figure \ref{fig:subham}). 
 
\bigskip

The restriction of the action of $A$ on the vertices labeled by these $d$-tuples, i.e. $\pi_{SV_L}\ A\ \pi_{SV_L}$, corresponds to the action of the adjacency matrix of a Hamming graph associated to tuples of length $L$. We want to diagonalize the chopped correlation matrix for this choice of subsystem. The projection operator onto $SV_L$ can be expressed as
\begin{align}
    \pi_{SV_L} = \underbrace{\ket{0}\bra{0} \otimes \ket{0}\bra{0}\otimes \dots \otimes  \ket{0}\bra{0}}_{d-L \text{ times}} \otimes \underbrace{\mathds{1}_{q \times q} \otimes \dots \otimes \mathds{1}_{q \times q}}_{L \text{ times}}.
\end{align}
\noindent Using \eqref{projk}, it is also possible to give an explicit expression for the projection operator onto the energy eigenspaces contained in the ground state $\pi_{SE} = \sum_{k\in SE} E_{d-k} $:
\begin{equation}
    \begin{split}
            \pi_{SE} = \sum_{k\in SE}\sum_{\substack{b \in F_2^d, \\ \text{wt}(b) = d-k}} {(b_1\mathds{1} + (-1)^{b_1}\frac{J_{q \times q}}{q})\otimes \dots \otimes(b_{d}\mathds{1} + (-1)^{b_{d}} \frac{J_{q \times q}}{q})} \ .
    \end{split}
\end{equation}

\noindent From these expressions, one can deduce the general form of the eigenvectors of $\pi_{SV_L} \pi_{SE} \pi_{SV_L}$. Indeed, let us recall that $\{\ket{\theta_i}\}$ with $i\in \{1,2,\dots,q\}$ is a set of eigenvectors of $J_{q\times q}$ and let us note that
\begin{align}
    \bra{0} (b_i\mathds{1} + (-1)^{b_i}\frac{J_{q \times q}}{q}) \ket{0} =
     \left\{
    \begin{array}{ll}
    	 \frac{1}{q} & \mbox{if } b_i = 0,  \\
    	\frac{q-1}{q} & \mbox{if } b_i = 1,
    \end{array}
\right. 
\end{align}
\noindent and that 
\begin{align}
    \ket{\theta_q} \bra{\theta_q} =   \frac{J_{q \times q}}{q} \quad \quad \text{and} \quad \quad \sum_{i = 1}^{q-1} \ket{\theta_i} \bra{\theta_i} = \mathds{1} - \frac{J_{q \times q}}{q}.
\end{align}
\noindent Thus, if we define the following vectors in $ \pi_{SV_L}(\mathbb{C}^{q})^{\otimes d}$:
\begin{align}
    \ket{\theta_{i_1}\dots \theta_{i_{L}} } = \underbrace{\ket{0} \otimes \ket{0}\otimes \dots \otimes  \ket{0} }_{d-L \text{ times}}\otimes \ket{\theta_{i_1}} \otimes \dots \otimes \ket{ \theta_{i_{L}}},
\end{align}
\noindent we can see that
\begin{multline}
    \pi_{SV_L} E_{d-k} \pi_{SV_L} \ket{\theta_{i_1}\dots \theta_{i_{L}} }   = \pi_{SV_L} E_{d-k}  \ket{\theta_{i_1}\dots \theta_{i_{L}} } \\
    =  
    \left\{
    \begin{array}{ll}
    	(\frac{1}{q})^{k-N_q}(\frac{q-1}{q})^{d-L-k+N_q} \binom{d-L}{k-N_q}\ket{\theta_{i_1}\dots \theta_{i_{L}} } & \mbox{if } d-L \geq k-N_q \geq 0,  \\
    	0 & \mbox{otherwise,}
    \end{array}
\right. 
\end{multline}
\noindent where $ N_q \equiv$ number of $\ket{\theta_q}$ in $\ket{\theta_{i_1}\dots \theta_{i_{L}} }$. Therefore, the action of the chopped correlation matrix on these vectors is
\begin{equation}
    \begin{split}
          C & \ket{\theta_{i_1}\dots \theta_{i_{L}} } =\\
          &= \Big[\sum_{\substack{ k \in SE \\ d-L \geq k - N_q \geq 0 }} 	\left(\frac{1}{q}\right)^{k-N_q}\left(\frac{q-1}{q}\right)^{d-L-k +N_q} \binom{d-L}{k-N_q}\Big] \ket{\theta_{i_1}\dots \theta_{i_{L}} }.
    \end{split}
\end{equation}
\noindent For all $N_q \in \{0,1,\dots,L\}$, we thus have an eigenvalue $\lambda_{N_q}$ of $C$, given by the following expression:
\begin{align}
    \lambda_{N_q} = \sum_{\substack{ k \in SE \\ d-L+ N_q \geq k \geq N_q }} 	\left(\frac{1}{q}\right)^{ k - N_q}\left(\frac{q-1}{q}\right)^{d-L- k + N_q} \binom{d-L}{ k - N_q}.
    \label{eigC}
\end{align}
\noindent As for its degeneracy $\binom{L}{N_q}(q-1)^{L-N_q}$, it corresponds to the number of vectors $\ket{\theta_{i_1}\dots\theta_{i_L}}$ having the same $N_q$. This result can then be used as an input in \eqref{entropy} to obtain the entanglement entropy $S$. 

\bigskip

\begin{figure}
\begin{subfigure}{.5\textwidth}
  \centering
  \includegraphics[scale = 0.4]{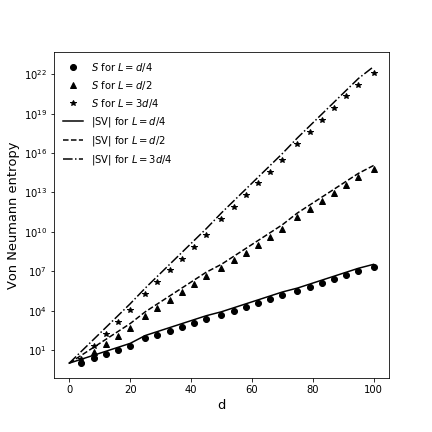}
  \caption{}
  \label{fig:sfig1}
\end{subfigure}%
\begin{subfigure}{.5\textwidth}
  \centering
  \includegraphics[scale = 0.4]{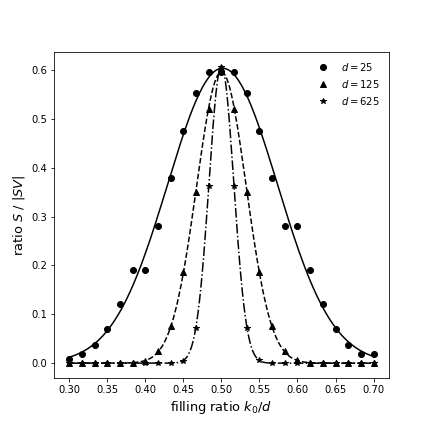}
  \caption{}
  \label{fig:sfig2}
\end{subfigure}
\caption{Entanglement entropy for cubes of dimension $L$ in cubes of dimension $d$. (a): entropy at $k_0 = d/2$ (half-filling) for sub-cubes of dimension $L = d/4$, $L = d/2$ and $L= 3d/4$. (b) Ratio of the entropy over the dimension $2^L$ of the sub-cubes ($L = d/4$) for different filling ratios of the Fermi sea. }
\label{fig:fig}
\end{figure}

In order to provide an explicit example, let us restrict ourselves to the case $SE=\{0,1,\dots ,k_0\}$. When $N_q > k_0$, we notice that $SE \cap \{N_q, N_q +1 ,\dots, d-L+N_q\}$ is empty and so $\lambda_{N_q}$ is $0$. Moreover, formula \eqref{eigC} corresponds to a sum over all the terms in a binomial distribution when $d-L+ N_q < k_0$. This implies that $\lambda_{N_q} = 1$. One can check that in these two cases, $\lambda_{N_q}$ does not contribute to \eqref{entropy}. We are left with $d-L+1$ eigenvalues to consider and obtain:
\begin{align}
    S = - \sum_{i =0}^{d-L} \binom{L}{d-k_0 - i}(q-1)^{d-k_0 -i}\left[F(i) \ln{F(i)} + (1 - F(i)) \ln{(1- F(i))}\right],
\end{align}
\noindent where $F$ is the cumulative binomial distribution function
\begin{align}
    F(i) = F(i;d-L,1/q) = \sum_{i' = 0}^i \binom{d-L}{i'}\left(\frac{1}{q}\right)^{i'}\left(\frac{q-1}{q}\right)^{d-L-i'}.
\end{align}
\noindent In terms of Krawtchouk polynomials, we also have
\begin{align}
    F(i) = 1 - \left(\frac{1}{q}\right)^{i+1}\left(\frac{q-1}{q}\right)^{d-L-i-1} \binom{d-L}{i+1}K_{d-L-i-1}(-1 ; 1-q, i+2).
\end{align}
\noindent In Figure \ref{fig:fig}, the entropy of sub-cubes in hypercubes are presented for different size ratios $L/d$ and filling ratios $k_0/d$. The left figure suggests that the entropy of Hamming subgraphs grows with their volume $|SV| = 2^L$, so that their state are nearly perfectly mixed. This does not contradict the area law, since every site in a Hamming subgraph has a least one neighbor outside the subsystem and is thus on the boundary. The figure on the right shows that $S$ peaks at half-filling $k_0/d$.

\section{Entanglement entropy for neighborhoods}
\label{s4}
Let us pick a reference tuple $v_0$ and consider all the sites labeled by tuples $v$ at a common distance $i$ from $v_0$. This set of sites is called the $i^{\text{th}}$ neighborhood of $v_0$. In this section, we want to find the entanglement entropy of subsystems which are bundles of neighborhoods of $v_0$. For a subset of distances $SD \subset \{0,1,\dots,d\}$, we take the subsystem 1 to be the set of sites $v$ for which the distance from $v_0$ is in $SD$, i.e. 
\begin{align}
    SV = \{v \in F^d_q \ : \ \partial(v_0,v) \in SD\}.
\end{align}

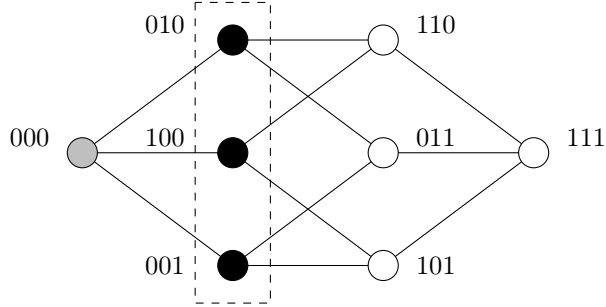
\begin{figure}[h]
    \centering
    \begin{tikzpicture}
\draw (-3,0) -- (-1,1.5);
\draw (-3,0) -- (-1,0);
\draw (-3,0) -- (-1,-1.5);

\draw (3,0) -- (1,1.5);
\draw (3,0) -- (1,0);
\draw (3,0) -- (1,-1.5);

\draw (-1,1.5) -- (1,1.5);
\draw (-1,1.5) -- (1,0);

\draw (-1,-1.5) -- (1,0);
\draw (-1,-1.5) -- (1,-1.5);

\draw (-1,0) -- (1,1.5);
\draw (-1,0) -- (1,-1.5);

\draw[black,fill=white] (1,1.5) circle (0.2cm);
\node[] at (1.7,1.7) {110};
\draw[black,fill=white] (1,0) circle (0.2cm);
\node[] at (1.7,0.2) {011};
\draw[black,fill=white] (1,-1.5) circle (0.2cm);
\node[] at (1.7,-1.5) {101};

\draw[black,fill=black] (-1,1.5) circle (0.2cm);
\node[] at (-1.9,1.7) {010};
\draw[black,fill=black] (-1,0) circle (0.2cm);
\node[] at (-1.9,0.2) {100};
\draw[black,fill=black] (-1,-1.5) circle (0.2cm);
\node[] at (-1.9,-1.5) {001};

\draw[black,fill=white] (3,0) circle (0.2cm);
\node[] at (3.7,0.2) {111};
\draw[black,fill=lightgray] (-3,0) circle (0.2cm);
\node[] at (-3.7,0.2) {000};

\draw[black, dashed] (-1.5, 2) rectangle (-0.5, -2);
\end{tikzpicture}

    \caption{A three dimensional cube and, in black, the $1^{\text{st}}$ neighborhood of the vertex $(0,0,0)$ (in gray).}
    \label{fig:ohye}
\end{figure}

\noindent For instance, we could consider $H(3,2)$ with $v_0  = (0,0,0)$ and $SD = \{1\}$. Then, the subsystem would be the sites labeled by $(0,0,1)$, $(0,1,0)$ or $(1,0,0)$ (see Figure \ref{fig:ohye}).

\newpage

\noindent Now, without loss of generality, we take $v_0 = (0,0,0,\dots,0)$. The projection operator onto the space associated to vertices at a distance $i$ from $v_0$ is defined as
\begin{align}
    E_i^* = \sum_{\substack{v \in F_q^d \\ \partial(v_0,v) = i}} \ket{v} \bra{v}.
    \label{projspace}
\end{align}
\noindent The projector over the subsystem is
\begin{align}
    \pi_{SV} &= \sum_{\substack{v \in F_q^d \\ \partial(v_0,v) \in SD}} \ket{v} \bra{v} = \sum_{i \in SD} E_i^* .
\end{align}
\noindent Thus, the chopped correlation matrix is
\begin{align}
    C = \sum_{i,i' \in SD} \sum_{k \in SE} E_i^*E_{d-k} E_{i'}^* \ .
    \label{choppy}
\end{align}
\noindent To diagonalize this chopped correlation matrix, we can take advantage of the fact that $C$ is expressed only in terms of elements in the Terwilliger algebra of the Hamming scheme (denoted by $\mathcal{T}$). This means that the vector space $(\mathbb{C}^{q})^{\otimes d}$ on which the operators $E_k$ and $E^*_i$ are acting corresponds to a $\mathcal{T}-$module and can be decomposed into irreducible submodules. This decomposition simplifies the problem, since it allows to work with one irreducible module at a time. In the next subsections, we give a quick overview of $\mathcal{T}$ and show how the decomposition in irreducible components of the representation on $(\mathbb{C}^q)^{\otimes d}$ is carried out.

\subsection{The Terwilliger algebra of the Hamming scheme}

Since the Hamming graphs are distance-regular, the following relations between adjacency matrices and projection operators are verified:
    \begin{itemize}
        \item $A_0 = \mathds{1}$ and $E_0 = J_{q^d \times q^d}/q^d$ ;
        \item $\sum_{i =0}^d A_i  = J_{q^d \times q^d}$ and $\sum_{i =0}^d E_i  = \mathds{1}$ ;
        \item $A_i \circ A_j = \delta_{ij} A_i$ and $E_i E_j = \delta_{ij} E_i$ ; 
        \item $A_i A_j = \sum_{k=0}^d p_{ij}^k A_k$ and $E_i \circ E_j = \frac{1}{q^d}\sum_{k=0}^d q_{ij}^k E_k$,
    \end{itemize}
\noindent where $(A\circ B)_{ij} = A_{ij}B_{ij} $ is the entry-wise product. The commuting algebra spanned by the adjacency matrices $A_i$ is in the present case the Bose-Mesner algebra of the Hamming scheme. It is important to note that the set of primitive idempotents $E_i$ also offers a basis for this algebra. Indeed, there exists coefficients $p_i(j)$ and $q_i(j)$ such that
\begin{align}
    A_i = \sum_{j=0}^d p_i(j) E_j \quad \quad \text{and} \quad \quad E_i = \frac{1}{|X|} \sum_{j = 0}^d q_i(j) A_j ,
\end{align}
\noindent where $|X|$ is the dimension of the matrices and is equal to $q^d$ for the Hamming scheme. From there, we can choose a reference vertex $v_0$ and introduce dual matrices $A^*_i(v_0)$ and $E_i^*(v_0)$ The dual adjacency matrices and dual primitive idempotents of the Hamming scheme are hence
\begin{align}
    [A_i^*(v_0)]_{vv} = q^d [E_i]_{v_0 v} \quad \quad \text{and} \quad \quad [E_i^*(v_0)]_{vv} = [A_i]_{v_0v} .
\end{align}
\noindent In the following, we use a simplified notation: $A_i^* =A^*_i(v_0)$ and $E_i^*=E_i^*(v_0)$. These matrices obey relations similar to the ones of their counterpart $A_i$ and $E_i$:
\begin{align}
    A^*_i A^*_j = \sum_{k=0}^d q_{ij}^k A^*_k \quad \quad \text{and} \quad \quad E^*_i E^*_j = \delta_{ij} E^*_i.
\end{align}
One can check that the definition of $E_i^*$ given above is equivalent to \eqref{projspace}. In the Hamming scheme, $A^*$ is given by the following tensor products:
\begin{align}
    A^* =  q \sum_{i = 1}^d \big(\underbrace{\mathds{1} \otimes \dots \otimes \mathds{1}}_{i-1 \text{ times}} \otimes \ \ket{0}\bra{0} \otimes \mathds{1} \otimes \dots \otimes \mathds{1}\big)  - d.
    \label{dudu}
\end{align}
\noindent It is easy to check that it acts diagonally on the set of vectors $\ket{v}$, $v \in F^d_q$, associated to sites in the graph. 

\bigskip

The Terwilliger algebra is the algebra spanned by the set of adjacency matrices $\{A_0, A_1, \dots, A_d\}$ and the set of dual adjacency matrices $\{A_0^*, A_1^*, \dots, A_d^*\}$ \cite{thini, thinii, thiniii}. In terms of the primitive idempotents, it can also be thought of as the algebra spanned by $\{E_0, E_1, \dots, E_d\}$ and $\{E_0^*, E_1^*, \dots, E_d^*\}$. Looking back at the definition of the chopped correlation matrix \eqref{choppy}, we see that $C$ is indeed part of the Terwilliger algebra of the Hamming scheme.

\subsection{Extracting the irreducible modules}

We now show how to decompose $(\mathbb{C}^{q})^{\otimes d}$ into the irreducible modules of the Terwilliger algebra of the Hamming scheme. This will allow to diagonalize $C$ one irreducible module at a time. 

\bigskip

This decomposition has been studied in the past \cite{GO2002399,LEVSTEIN20061}. It has been applied to the computation of entanglement entropy of coupled harmonic oscillators on Hamming networks in \cite{jafarizadeh2014entanglement,Jafarizadeh2015EntanglementEI} where the Schmidt numbers and thus the entropy could be extracted from the potential matrix of the system expressed in the basis associated to the decomposition. 

\subsubsection{Irreducible representations for the hypercube $H(d,2)$}

First, we consider the case $H(d,2)$, i.e. the Terwilliger algebra of hypercubes. We see then from \eqref{fia} and \eqref{dudu} that the adjacency and dual adjacency matrices are given in terms of Pauli matrices:
\begin{align}
    A =  \sum_{i = 1}^d \big(\underbrace{\mathds{1} \otimes \dots \otimes \mathds{1}}_{i-1 \text{ times}} \otimes \ \sigma^x \otimes \mathds{1} \otimes \dots \otimes \mathds{1}\big)
    \label{dudu}
\end{align}
and
\begin{align}
    A^* =  \sum_{i = 1}^d \big(\underbrace{\mathds{1} \otimes \dots \otimes \mathds{1}}_{i-1 \text{ times}} \otimes \ \sigma^z \otimes \mathds{1} \otimes \dots \otimes \mathds{1}\big) .
    \label{dudu}
\end{align}
\noindent $A$ and $A^*$ can thus be interpreted in terms of elements in the representation of $\mathfrak{su}(2)$ constructed from the tensor product of $d$ irreducible representations of spin $1/2$. Indeed, if $s_i^x$, $s_i^y$ and $s_i^z$ are the following operators:
\begin{align}
     s_i^j = \underbrace{\mathds{1} \otimes \dots \otimes \mathds{1}}_{i-1 \text{ times}} \otimes \ \frac{\sigma^j}{2} \otimes \mathds{1} \otimes \dots \otimes \mathds{1}, \quad \quad \text{with } j \in \{x,y,z\},
\end{align}
\noindent  giving for each $i$ a two dimensional representation of $\mathfrak{su}(2)$, it follows that the adjacency and dual adjacency matrices are proportional to the components $s^x$ and $s^z$ of the total spin operator, i.e.
\begin{align}
    A = 2 \sum_{i = 1}^d s_i^x = 2 s^x \quad \quad \text{and} \quad \quad  A^* = 2 \sum_{i = 1}^d s_i^z = 2 s^z.
\end{align}
\noindent From this, one deduces that the irreducible modules of the Terwilliger algebra of $H(d,2)$ are the vector spaces associated to the irreducible representations of $\mathfrak{su}(2)$ contained in the $d$-fold product of the fundamental representation of $\mathfrak{su}(2)$. For $d$ even (resp. odd) we know from the standard Clebsh-Gordan decomposition that for each $j$ in $\{ 0, 1,\dots, \frac{d}{2}\}$ (resp. in $\{ 1/2, 3/2, \dots, \frac{d}{2}\}$) there exists $\frac{2j + 1}{d+1}\binom{d+1}{\frac{d}{2} - j}$ subspaces of $(\mathbb{C}^2)^{\otimes d}$ spanned by vectors $\{\ket{j,m}\}_{-j \leq m \leq j}$ such that
\begin{equation}
    \begin{split}
            A \ket{j,m} &= \sqrt{(j+m +1)(j-m)} \ket{j,m+1}  \\ & \quad + \sqrt{(j-m +1)(j+m)} \ket{j,m-1}
    \end{split}
\end{equation}
\noindent and
\begin{equation}
    \begin{split}
            A^* \ket{j,m} &= 2m \ket{j,m}
    \end{split}
\end{equation}
\noindent The label necessary to distinguish the different spaces of dimension $2j +1$ is kept implicit. It is also important to note that since the $\ket{j,m}$ are eigenvectors of $A^*$ with eigenvalue $2m$, they are contained in the $(\frac{d}{2} - m)$-th neighborhood of the hypercube, i.e. they are linear combinaisons of vectors $\ket{v}$ of weight $\text{wt}(v) = \frac{d}{2} - m$. 

\subsubsection{Irreducible representations for the Hamming graph $H(d,q)$}

Now that we have the decomposition for $q = 2$, it can be used to obtain a similar result for a general Hamming graph $H(d,q)$. More precisely, we want to identify subspaces of $(\mathbb{C}^q)^{\otimes d}$ on which the action of $A$ and $A^*$ is similar to their action in the case of a hypercube. First, let us define a set of vectors $\{\ket{\tilde{n}}\}_{0 \leq n < q}$ which gives a basis of $\mathbb{C}^q$:
\begin{align}
    \ket{\tilde{0}} = \ket{0}, \quad \quad \quad \ket{\tilde{1}} = \frac{1}{\sqrt{q-1}}\sum_{i = 1}^{q-1}\ket{i};
\end{align}
\noindent and for $n > 1$, the degeneracy of the eigenspace of $J$ associated to the eigenvalue $0$ allows to take them such that 
\begin{align}
    \bra{\tilde{n}}\ket{\tilde{m}} = \delta_{mn}, \quad \bra{0}\ket{\tilde{n}} = 0 \quad \text{and} \quad J\ket{\tilde{n}} = 0.
\end{align}
\noindent When $q=2$, we notice that $\ket{\tilde{1}} = \ket{1}$ and that this basis matches with the usual one. Futhermore, for any $v \in F^d_q$, we can define $\ket{\tilde{v}} = \ket{\tilde{v_1}} \otimes \ket{\tilde{v_2}} \otimes \dots \otimes \ket{\tilde{v_d}}$, which gives a new basis for $(\mathbb{C}^{q})^{\otimes d}$. 

\bigskip

Let us divide this basis of $(\mathbb{C}^{q})^{\otimes d}$ in subsets. For two $d$-tuples $v$ and $v'$ in $F^d_q$, we say that $v \sim v'$ if $\forall \ i \in \{1, \dots, d\}$ we either have (1) $v_i = v_i'$, (2) $v_i = 0$ and $v_i' = 1$ or (3) $v_i' = 0$ and $v_i = 1$. In other words, we say that two tuples are equivalent if they differ only on their binary part. We refer to the set of $d$-tuples equivalent to $v$ as $[v]$. We can define the subspace of $(\mathbb{C}^{q})^{\otimes d}$ generated by all the vectors associated to tuples in $[v]$ as $V_{[v]} = \text{span}\{\ket{\tilde{v}'}\}_{v' \in [v]}$. Thus, we have the following decomposition of our vector space:
\begin{align}
    (\mathbb{C}^{q})^{\otimes d} = \bigoplus_{v\  \in\  F^d_q/\sim } V_{[v]}.
\end{align}
\noindent We are interested in characterizing the action of $A$ on a subspace $V_{[v]}$. With no loss of generality, we pick a tuple $v$ for which the first $n$ elements are binary. Since $d$-tuples in $[v]$ differ only at positions where they take the value $0$ or $1$, we can refer to the basis vectors $\ket{\tilde{v}'}$ of $V_{[v]}$ with a binary $n$-tuples $b = (b_1, \dots, b_{n})$. Indeed, we have
\begin{equation}
\begin{split}
    \ket{\tilde{v}'} = \underbrace{\ket{\tilde{v}'_1} \otimes \dots \otimes \ket{\tilde{v}'_{n}} }_{\text{binary part}} \otimes\underbrace{ \ket{\tilde{v}'_{n+1}} \otimes \dots \otimes \ket{\tilde{v}'_{d}}}_{\text{fixed by the choice of }[v]}\\
    = \ket{\tilde{b_1}} \otimes \dots \otimes \ket{\tilde{b}_{n}} \otimes \ket{\psi} \equiv \ket{b},
    \end{split}
\end{equation}
\noindent where $\ket{\psi} = \ket{\tilde{v}'_{n+1}} \otimes \dots \otimes \ket{\tilde{v}'_{d}}$ is the fixed part. To find the action of $A$ on a vector $\ket{b}$, we notice that
\begin{align}
    (J - \mathds{1}) \ket{\tilde{0}} = \sqrt{q-1} \ket{\tilde{1}}, \quad \quad (J - \mathds{1}) \ket{\tilde{1}} = \sqrt{q-1} \ket{\tilde{0}} + (q-2)\ket{1},
\end{align}
\noindent and that for $n > 1$
\begin{align}
(J - \mathds{1}) \ket{\tilde{n}} = - \ket{\tilde{n}}.
\end{align}
\noindent Therefore, we can express the action of $A$ in terms of tensor products of Pauli matrices in the following way:
\begin{equation}
    \begin{split}
    A \ket{b} &= \Big[\sqrt{q-1} \sum_{i = 1}^n \big(\underbrace{\mathds{1} \otimes \dots \otimes \mathds{1}}_{i-1 \text{ times}} \otimes \ \sigma^x \otimes \mathds{1} \otimes \dots \otimes \mathds{1}\big) \ket{\tilde{b_1}} \otimes \dots \otimes \ket{\tilde{b_n}} \Big] \otimes \ket{\psi}\\
    & - \Big[\frac{(q-2)}{2} \sum_{i = 1}^n \big(\underbrace{\mathds{1} \otimes \dots \otimes \mathds{1}}_{i-1 \text{ times}} \otimes \ \sigma^z \otimes \mathds{1} \otimes \dots \otimes \mathds{1}\big) \ket{\tilde{b_1}} \otimes \dots \otimes \ket{\tilde{b_n}} \Big] \otimes \ket{\psi} \\
    &+ \left(\frac{qn}{2}-d\right)\ket{b}, 
    \end{split}
\end{equation}
\noindent where 
\begin{align}
    \sigma^x \ket{\tilde{0}} =  \ket{\tilde{1}}, \quad \sigma^x \ket{\tilde{1}} =  \ket{\tilde{0}}, \quad \sigma^z \ket{\tilde{0}} =  \ket{\tilde{0}} \quad \text{and} \quad \sigma^z \ket{\tilde{1}} = -\ket{\tilde{1}}.
\end{align}
\noindent For the action of $A^*$ on $V_{[v]}$, we see from \eqref{dudu} that:
\begin{align} 
    A^* \ket{b} = q (n- \text{wt}(b)) - d \ket{b} .
\end{align}
\noindent At this point, we can apply the results obtained for $q=2$. Thus, given a subspace $V_{[v]}$ characterized by a $n$ even (resp. odd), we know that for each $j$ in $\{ 0, 1,\dots, \frac{n}{2}\}$ (resp. in $\{ 1/2, 3/2, \dots, \frac{n}{2}\}$) there exists $\frac{2j + 1}{n+1}\binom{n+1}{\frac{n}{2} - j}$ subspaces spanned by vectors $\{\ket{j,m}\}_{-j \leq m \leq j}$ such that
\begin{equation}
    \begin{split}
    A\ket{j,m} &=  \sqrt{(q-1)(j+m+1)(j-m)}\ket{j,m+1} \\
    & + \sqrt{(q-1)(j+m)(j-m+1)}\ket{j,m-1} \\
    &+ (\frac{qn}{2} - d - (q-2)m ) \ket{j,m}
    \end{split}
\end{equation}
\noindent and
\begin{align}
    A^* \ket{j,m} =  qm + \frac{nq}{2} - d \ket{j,m}.
\end{align}
\noindent Let $\ell$ label the degenerate subspaces of dimension $2j + 1$. We see that $K_{j,\ell,[v]} =  \text{span}\{\ket{j,m} : m \in \{-j, \dots j\}\}$ corresponds to the irreducible $\mathcal{T}$-submodules. Therefore, the decomposition needed is:
\begin{align}
    (\mathbb{C}^{q})^{\otimes d} = \bigoplus_{v \in  F^d_q/\sim  } \bigoplus_{j,\ell} K_{j,\ell,[v]}.
    \label{dec}
\end{align}
\noindent For the general Hamming case, it is worth noting that $\ket{j,m}$ is in the $(d-\frac{n}{2} - m)$-th neighborhood of the graph and that each irreducible module $K_{j,\ell,[v]}$ contains no more than one vector per neighborhood. In algebraic terms, this amounts to say that $\mathcal{T}$ is \textit{thin} and it is well known that the Terwilliger algbera of the Hamming scheme has this property \cite{thiniii}. 

\subsubsection{Chopped correlation marix for neighborhoods and entanglement entropy}

Now, let us see how \eqref{dec} simplifies the diagonalization of the chopped correlation matrix. First, we can consider
\begin{align}
    \underbar{c}^\dagger_{j,\ell,m,v} = \sum_{v' \in  F^d_q  } \bra{v'}\ket{j,m} c_{v'}^\dagger \quad \text{and} \quad     \underbar{c}_{j,\ell,m,v} = \sum_{v' \in  F^d_q  } \bra{j,m}\ket{v'} c_{v'}.
\end{align}
\noindent One can check that these fermionic operators respect the canonical anticommutation relations. Expressing the Hamiltonian \eqref{hamil2} in terms of these operators, one finds that it decomposes into a sum of Hamiltonians $\widehat{\mathcal{H}}_{K_{j,\ell,[v]}}$ acting independently on orthogonal subspaces:
\begin{equation}
    \begin{split}
    \widehat{\mathcal{H}} &= \sum_{v \in F^d_q/\sim } \sum_{j,\ell} \sum_{m,m'=-j}^{j} \bra{j,m}[\sum_{i}\alpha_i A_i]\ket{j,m'} \underbar{c}_{j,\ell,m,v}^\dagger\underbar{c}_{j,\ell,m',v}\\
    &= \sum_{v \in  F^d_q/\sim} \sum_{j,\ell} \widehat{\mathcal{H}}_{K_{j,\ell,[v]}}\ .
    \end{split}
\end{equation}
\noindent Since each module contains at most one vector per neighborhood, the operators $\widehat{\mathcal{H}}_{K_{j,\ell,[v]}}$ can be interpreted as Hamiltonians of free fermions on chains. Similarly for $C$, we have
\begin{equation}
    \begin{split}
    C &= \pi_{SV}\  \pi_{SE}\ \pi_{SV}\\
    &= \sum_{v \in F^d_q/\sim } \sum_{j,\ell} [\pi_{SV} \pi_{SE} \pi_{SV}]_{K_{j,\ell,[v]}} \\
    &= \sum_{v \in F^d_q/\sim } \sum_{j,\ell} [C]_{K_{j,\ell,[v]}} \ ,
    \end{split}
\end{equation}
\noindent where $[C]_{K_{j,\ell,[v]}}$ is the restriction of the chopped correlation matrix to the subspace $K_{j,\ell,[v]}$. The entries of $[C]_{K_{j,\ell,[v]}}$ can be obtained in the following way. For $\ket{j,\omega_{k'}}$ an eigenvector of $[A]_{K_{j,\ell,[v]}}$, we get a three terms recurrence relations for the overlap coefficients $Q_{m,k'} = \bra{j,\omega_{k'}}\ket{j,m}$ by considering the matrix element $\bra{j,\omega_{k'}}[A]_{K_{j,\ell,[v]}}\ket{j,m}$. We have
\begin{equation}
    \begin{split}
    \omega_{k'} Q_{m, k'} &= \sqrt{(q-1)(j+m+1)(j-m)}Q_{m+1,k'} + \left(\frac{nq}{2}-d - (q-2)m\right)Q_{m,k'} \\
    & + \sqrt{(q-1)(j-m+1)(j+m)}Q_{m-1,k'}\ .
    \end{split}
\end{equation}
\noindent This leads to the identifications $\omega_{k'} = q(j-k' + \frac{n}{2}) - d$ and to
\begin{align}
    Q_{m,k'} = \sqrt{\binom{2j}{k'}}\sqrt{\binom{2j}{j-m}}q^{-j}(q-1)^{\frac{j-m+k'}{2}}K_{j-m}(k' ; (q-1)/q , 2j ) \ .
\end{align}
\noindent Therefore, we find for $m,m' \in SD\cap\{-j, \dots, j\}$ that
\begin{align}
    \bra{j,m}[C]_{K_{j,\ell,[v]}}\ket{j,m'}= \sum_{j-k'+\frac{n}{2} \in SE} Q_{m,k'} Q_{m',k'} \ .
\end{align}
\noindent Knowing all the entries of the matrix $[C]_{K_{j,\ell,[v]}}$, we can extract its eigenvalues, i.e. find the zeros of its characteristic polynomial of degree $| SD\cap\{d - \frac{n}{2} - j, \dots, d - \frac{n}{2} + j\}| $.  This degree is at most the number of neighborhoods in the subsystem. It is in general significantly less than the dimension of $C$, which is equal to the number of sites in all the neighborhoods in the subsystem. In particular, when we consider a single neighborhood, this decomposition actually yields the diagonalization. Indeed, $ [C]_{K_{j,\ell,[v]}}$ is then a $1$-dimensional matrix for which the only entry is its eigenvalue. If $SV$ is the $i^{\text{th}}$ neighborhood, the eigenvalues of $C$ are
\begin{equation}
    \begin{split}
         \lambda_{n,j} &= \sum_{k \in SE} Q_{d-\frac{n}{2}-i,\ j-k + \frac{n}{2}} ^2.
    \end{split}
\end{equation}
\noindent The degeneracy is given by the product of the number of equivalence classes $[v]$ with $d$-tuples containing $n$ binary terms and the number of representations of $\mathfrak{su}(2)$ of dimension $2j+1$ in the hypercube $H(n,2)$. Thus, it is given by  $D_{n,j} = (q-2)^{d-n}\frac{2j + 1}{d+1}\binom{d+1}{\frac{d}{2} - j}\binom{d}{n}$.
\begin{figure}[h]
\begin{subfigure}{.5\textwidth}
  \centering
  \includegraphics[scale = 0.4]{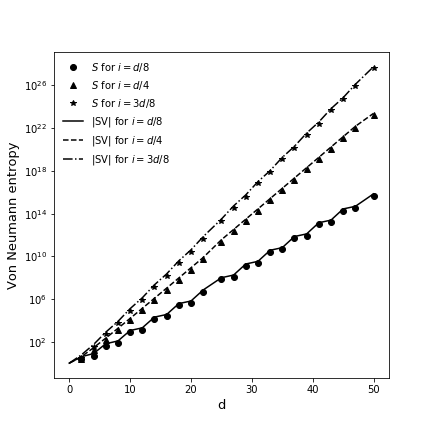}
  \caption{}
  \label{fig:sfig1}
\end{subfigure}%
\begin{subfigure}{.5\textwidth}
  \centering
  \includegraphics[scale = 0.4]{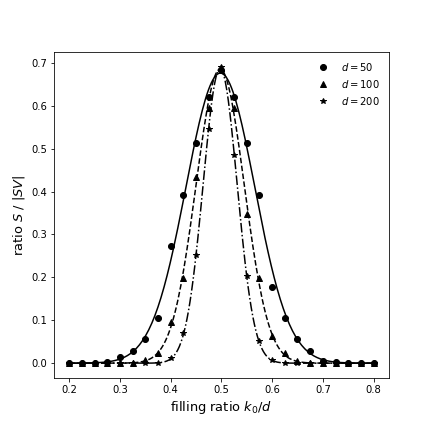}
  \caption{}
  \label{fig:sfig2}
\end{subfigure}
\caption{Entanglement entropy for neighborhoods of $d$-cubes. (a): entropy at $k_0 = d/2$ (half-filling) for the neighborhoods $i = d/8$, $i = d/4$ and $i = 3d/8$. (b) Ratio of the entropy over the dimension of the neighborhood $i = d/4$ for different filling ratios of the Fermi sea. }
\label{fig:fig2}
\end{figure}
In Figure \ref{fig:fig2}, the entropy of neighborhoods of hypercubes are presented for different distance ratios $i/d$ and filling ratios $k_0/d$. The left figure suggests that the entropy of a neighborhood grows with its volume $|SV|$ and thus that their state are nearly perfectly mixed. The figure on the right shows that $S$ peaks at half-filling $k_0/d$. We recall that similar results were obtained for Hamming subgraphs.
\newpage
\subsubsection{Interlude : $\mathcal{T}$ as a quotient of the Onsager algebra}

Before considering the case where $SV$ is made of a large number of neighborhoods, let us make an additional remark about $\mathcal{T}$. It is interesting to note that the decomposition of the vector space in $\mathcal{T}$-submodules has brought up a relation between the Terwilliger algebra of $H(d,q)$ and a quotient of the Onsager algebra $\mathcal{O}$ \cite{Baseilhac_2018,onsagerDavies,onsager,roan1991onsager}. To see this, we first highlight the fact that the relation between $\mathcal{T}$ and $\mathfrak{su}(2)$ extends to $q\neq2$. In terms of the matrices $s^x$ and $s^z$ from an irreducible representation of dimension $2j+1$, we note that
\begin{align}
    [A]_{K_{j,\ell,[v]}} = 2\sqrt{q-1} s^x - {(q-2)}s^z+ \left(\frac{nq}{2}-d\right)\mathds{1},
    \label{A1}
\end{align}
\noindent and
\begin{align}
    [A^*]_{K_{j,\ell,[v]}} = qs^z + \left(\frac{qn}{2}-d\right)\mathds{1}.
    \label{A2}
\end{align}
\noindent Using these expression and the commutation relations of $\mathfrak{su}(2)$, one can easily check that the dual and non-dual adjacency matrices of the Hamming scheme verify the following relations:
\begin{align}
    [A,[A,[A,A^*]]] = q^2 [A,A^*] \quad \quad \text{and} \quad \quad  [A^*,[A^*,[A^*,A]]] = q^2 [A^*,A].
\end{align}
\noindent These were first obtained by Terwilliger in \cite{thiniii}. Taking $\mathcal{A}_0 = \frac{4}{q} A$ and $\mathcal{A}_1 = \frac{4}{q} A^*$, one also sees that these operators verify the Dolan-Grady relations \cite{DolanGrady}:
\begin{align}
    [\mathcal{A}_0,[\mathcal{A}_0,[\mathcal{A}_0,\mathcal{A}_1]]] = 16 [\mathcal{A}_0,\mathcal{A}_1] \quad \text{and} \quad  [\mathcal{A}_1,[\mathcal{A}_1,[\mathcal{A}_1,\mathcal{A}_0]]] = 16 [\mathcal{A}_1,\mathcal{A}_0],
\end{align}
\noindent which serve as defining relations for the Onsager algebra $\mathcal{O}$. This algebra is in general of infinite dimension and the rest of its generators, $\mathcal{A}_n$ and $\mathcal{G}_n$ with $n \in \mathbb{Z}$, are defined by the equivalent presentation \cite{onsager}: 
\begin{align}
    [\mathcal{A}_n,\mathcal{A}_m] = 4 \mathcal{G}_{n-m},  \quad [\mathcal{G}_n,\mathcal{A}_m] = 2 \mathcal{A}_{n+m} - 2 \mathcal{A}_{m-n}  \quad \text{and} \quad [\mathcal{G}_n,\mathcal{G}_m] = 0.
\end{align}
\noindent We see that $\mathcal{T}$ gives a representation of $\mathcal{O}$. It is however of finite dimension. Indeed, one can check using \eqref{A1} and \eqref{A2} that the following Davies relations hold \cite{Davies_1990}:
\begin{align}
    \mathcal{A}_{p+2} - \mathcal{A}_{p-1} +\frac{q+4}{q} (\mathcal{A}_{p+1} - \mathcal{A}_{p} ) = 0
    \label{D1}
\end{align}
\noindent and
\begin{align}
    \mathcal{G}_{p+2} - \mathcal{G}_{p-1} +\frac{q+4}{q}  (\mathcal{G}_{p+1} - \mathcal{G}_{p} ) = 0,
    \label{D2}
\end{align}
\noindent for all $p \in \mathbb{Z}$. In other words, the generators can all be expressed as linear combinations of $\{\mathcal{A}_{-1},\mathcal{A}_0,\mathcal{A}_1, G_1\}$. Using the expression of these generators in terms of $\{s^x, s^y, s^z,\left(\frac{nq}{2}-d\right)\mathds{1} \}$, one can also check that they are linearly independent. Thus, $\mathcal{T}$ for $H(d,q)$ yields a Lie algebra isomorphic to the quotient of $\mathcal{O}$ by $\eqref{D1}$ and $\eqref{D2}$. 
 
\subsection{Entanglement entropy for multiple neighborhoods}
\label{s5}
\noindent The decomposition in irreducible modules has simplified the diagonalization of the chopped correlation matrix $C$ by allowing to work with matrices of lower dimension. However, we may still have to diagonalize large matrices with few non-zero entries and many eigenvalues approaching $0$ or $1$ when the subsystem is composed of many neighborhoods. This happens for instance if we take as subsystem 1 all the vertices at a distance lower than some large integer $N$ from a given site. Then,
\begin{align}
    \pi_{SV} = \sum_{i = 0}^N E_i^*.
\end{align}
\noindent In the spirit of the celebrated time and band limiting approach \cite{GB_2018,Landau1985,Slepian1983}, we would now wish to find a simple operator $T$ with practical diagonalization properties such that
\begin{align}
    [C, T] = 0
\end{align}
\noindent and that it thus shares with $C$ common eigenvectors.
\subsubsection{The generalized algebraic Heun operator}

\noindent Since we are working with a distance-regular graph, we can adopt the approach based on Heun operators developed in \cite{crampe2020entanglement, Cramp__2019,Cramp__2020} to construct $T$. This leads us to look at the most general symmetric block-tridiagonal operator in the Terwilliger algebra of the Hamming scheme:
\begin{align}
    T = \{A, A^*\} + \mu A^* + \nu A,
\end{align}
\noindent These are referred to as generalized algebraic Heun operators. We want to fix $\mu$ and $\nu$ to allow $T$ to commute with both $\pi_{SV}$ and $\pi_{SE}$, assuring that $T$
also commutes with $C$. For a given irreducible module $K_{j,\ell,[v]}$, the action of $T$ is tridiagonal on the eigenbases of both $A$ and $A^*$:
\begin{equation}
    \begin{split}
    T \ket{j,m} &= (\nu + 2qm + nq + q -2d)\sqrt{(q-1)(j+m+1)(j-m)}\ket{j,m+1} \\
    & + (\nu + 2qm + nq - q -2d)\sqrt{(q-1)(j+m)(j-m+1)}\ket{j,m-1} \\
    &+ \left[(\nu + 2qm + nq -2d)\left(\frac{qn}{2} - d - (q-2)m \right) +\mu\left(qm + \frac{nq}{2} -d\right)\right] \ket{j,m}
    \end{split}
    \label{T1}
\end{equation}
\noindent and 
\begin{equation}
    \begin{split}
          T \ket{j,\omega_{k'}} &= (\mu + \omega_{k'} + \omega_{k'+1})\sqrt{(q-1)(2j-k')(k'+1)}\ket{j,\omega_{k'+1}}  \quad \quad \quad \ \ \\
    & \quad +(\mu + \omega_{k'} + \omega_{k'-1})\sqrt{(q-1)(k')(2j-k'+1)}\ket{j,\omega_{k' - 1}} \\
    & \quad + \left[2\omega_{k'}\left(\frac{qn}{2} - d - (q-2)(j-k') \right) +\nu \omega_{k'}\right] \ket{j,\omega_{k'}},
    \end{split}
    \label{T2}
\end{equation}

\noindent where we recall that $\omega_{k'} = q(j-k' + \frac{n}{2}) - d$. As for the action of the projection operators on these vectors, they are given by
\begin{align}
\pi_{SV} \ket{j,m} = 
         \left\{
    \begin{array}{ll}
    	 \ket{j,m} & \mbox{if } d-\frac{n}{2} - m \leq N\\
    	0 & \mbox{otherwise }
    \end{array}
\right. 
\label{P1}
\end{align}
\noindent and
\begin{align}
\pi_{SE} \ket{j,\omega_{k'}} = 
         \left\{
    \begin{array}{ll}
    	 \ket{j,\omega_{k'}} & \mbox{if } j - k' +\frac{n}{2} \leq k_0\\
    	0 & \mbox{otherwise. }
    \end{array}
\right. 
\label{P2}
\end{align}
\noindent From \eqref{T1} and \eqref{P1}, we see that $[T,\pi_{SV}] = 0 $ if $(\nu + 2q(d-N) -q - 2d) = 0$.  Similarly, from \eqref{T2} and \eqref{P2}, we see that $[T,\pi_{SE}] = 0 $ if $(\mu + 2qk_0 - q -2d) = 0$. Thus, if we take
\begin{align}
    \nu = -2q(d-N) +q + 2d \quad \quad \text{and} \quad \quad \mu = -2qk_0 + q +2d,
\end{align}
\noindent we see that 
\begin{align}
    [T, C] = [T, \pi_{SV} \pi_{SE} \pi_{SV}] = 0.
    \label{commute}
\end{align}
\noindent Since the choice of $\mu$ and $\nu$ does not depend on $n$ or $j$, \eqref{commute} holds for all the irreducible modules. To diagonalize $T$, we recall \eqref{A1} and \eqref{A2} which imply
\begin{equation}
    \begin{split}
    [T]_{K_{j,\ell,[v]}} &= 2q \sqrt{q-1} \{s^x,s^z\} - q{(q-2)}\{s^z, s^z\}  +2 \sqrt{q-1}(2\left(\frac{nq}{2} - d\right) + \nu)s^x \\
    &+ (4\left(\frac{nq}{2} -d\right) + \mu q - \nu ({q-2}))s^z +  \left(\frac{nq}{2} - d\right)(2\left(\frac{nq}{2} - d\right) + \mu + \nu).
    \end{split}
    \label{Heunr}
\end{equation}

\begin{figure}[h]
\begin{subfigure}{.5\textwidth}
  \centering
  \includegraphics[scale = 0.4]{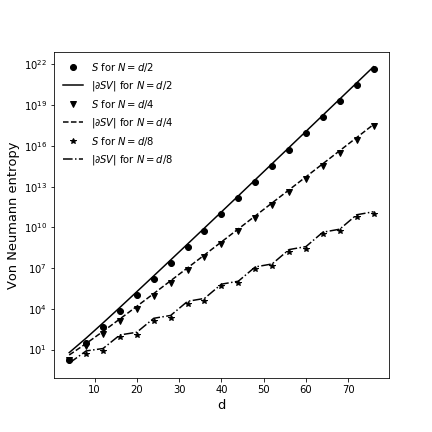}
  \caption{}
  \label{fig:sfig1}
\end{subfigure}%
\begin{subfigure}{.5\textwidth}
  \centering
  \includegraphics[scale = 0.4]{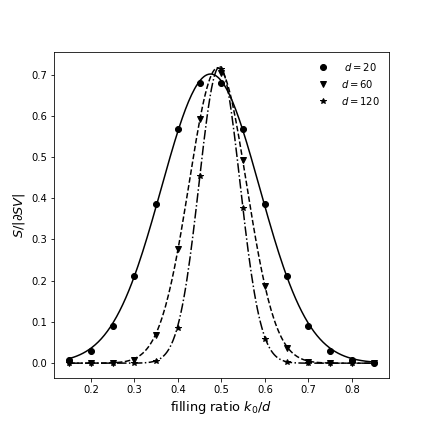}
  \caption{}
  \label{fig:sfig3}
\end{subfigure}
\caption{Entanglement entropy of the sites at a distance $\leq N$ from a fixed vertex. (a): entropy at $k_0 = d/2$ (half-filling) for $N = d/8$, $N = d/4$ and $N = d/2$. (b) Ratio of the entropy $S$ over $|\partial SV|$ (the number of sites at the boundary of the subsystem) for different filling ratios of the Fermi sea. }
\label{fig:fig3}
\end{figure}

\noindent This operator has a well behaved spectrum, making its eigenvectors easy to obtain numerically. Computing the entropy is then a matter of acting on these vectors with the chopped correlation matrix to extract its spectrum and applying \eqref{entropy}. Figure \ref{fig:fig3} presents results obtained with this method in the case of hypercubes containing up to $2^{120}$ sites. As expected, the entropy seems to verify an area law as we approach the thermodynamic limit.

\bigskip

The diagonalization of $T$ can also be approached analytically. Indeed, the operator \eqref{Heunr} corresponds to a $BC$-Gaudin magnet Hamiltonian in a magnetic field and can be diagonalized using the modified algebraic Bethe ansatz. This is briefly presented in the following subsection (for more details, see \cite{bernard2020heun, Crampe_2017}).

\subsubsection{The modified algebraic Bethe ansatz and the spectrum of $T$}

In a nutshell, the modified algebraic Bethe ansatz works in the following way. From a solution $r$ of the non-standard classical Yang-Baxter equation, one constructs a $K$- matrix which satisfies the classical reflection equation and defines a transfer matrix $t(u) = \text{tr}K(u)^2$. This matrix $t(u)$ satisfies $[t(u),t(v)]=0$, that is the matrices associated to different values of the parameter commute. It is known how to find solutions $r$ and $K$ for which $t(u)$ can be identified as a BC-Gaudin magnet Hamiltonian in a magnetic field (and thus, as $T$) when $u \rightarrow 0$. It is also known that such transfer matrix can be diagonalized by vectors of the following form:
\begin{align}
    \ket{\boldsymbol{z}} = B(z_1,1)B(z_2,2)\dots B(z_{2j+1},2j+1) \ket{j,-j},
    \label{antz}
\end{align}
\noindent where $B(z,n)$ are operators that can be written in terms of $\mathfrak{su}(2)$ generators and are constructed to satisfy particular commutation relations with the entries of the $K$-matrix. For the ansatz \eqref{antz} to be an eigenvector of $t(u)$, a set of equations depending on $\boldsymbol{z}$ needs to be verified. These are the Bethe equations that are given by:
\begin{equation}
\begin{split}
   \frac{e^{2i\theta} j ({z}_k^2 - 1)}{(e^{2i\theta} {z}_k - 1)(e^{2i\theta} - {z}_k)} + \frac{({z}_k^2 +1)(1 - \frac{\nu}{2})- \mu {z}_k}{({z}_k^2 - 1)  } +\sum_{\genfrac{}{}{0pt}{2}{p=1}{p\neq k}}^{\mathcal{M}} \frac{{z}_p({z}_k^2-1)}{({z}_k-{z}_p)({z}_k {z}_p-1)} =0,
\end{split}
 \label{BeqC} 
\end{equation}
\noindent where $\theta = \frac{1}{2}\arccot{\frac{\sqrt{q-1}}{q-2}}$ and the dimension of $\pi_{SV} K_{j,\ell,[v]} \pi_{SV}$ is $\mathcal{M} = N - d + j + \frac{n}{2}$. To a given solution $\bar{\boldsymbol{z}}$, there corresponds an eigenvector of $t(u)$ and thus of $T$. The associated eigenvalue of the generalized Heun operator is
\begin{equation}
 \ t_{\boldsymbol{\Bar{z}}} =  q \left(j\cos{(2 \theta)} + \frac{\mu}{2} - \frac{1}{2} \sum_{i=1}^\mathcal{\mathcal{M}} (\Bar{z}_i+ \frac{1}{\Bar{z}_i})\right) + q\left(\frac{nq}{2} -d \right)(nq - 2d + \mu + \nu).
\end{equation}

\subsubsection{The spectrum of the chopped correlation matrix}
Once the generalized Heun operator is diagonalized, we can use the results to recover the spectrum of $C$, which shares with $T$ the same eigenspaces. Usually, this is done by acting with $C$ on the eigenvectors of $T$ and by reading out the eigenvalues of the chopped correlation matrix from the outcomes. Here, we present an alternative way to recover the spectrum. We construct a polynomial $P$ of order $\mathcal{M}-1$ such that on $SV$
\begin{align}
    [C]_{K_{j,\ell,[v]}} = P([T]_{K_{j,\ell,[v]}}) = \sum_{i=0}^{\mathcal{M}-1} a_i [T]_{K_{j,\ell,[v]}}^i.
\end{align}
\noindent If we have $P$ and an eigenvalue $t_{\bar{\boldsymbol{z}}}$ of $[T]_{K_{j,\ell,[v]}}$, it is obvious that $P(t_{\bar{\boldsymbol{z}}})$ gives an eigenvalue of $C$. To prove the existence of $P$, it is sufficient to know that the Heun operator and the chopped correlation matrix commute and to notice that $[T]_{K_{j,\ell,[v]}}$ is irreducible tridiagonal on $SV$. Therefore, the first $\mathcal{M}$ powers of $T$ are linearly independent. The fact that $T$ is tridiagonal also allows to determine $P$. Indeed, since we have that
\begin{align}
    \beta_{i,j+m} \equiv \bra{j,-j} [T]_{K_{j,\ell,[v]}}^i \ket{j,m} = 0 \quad \quad \text{if } i < j+m,
\end{align}
\noindent we are led to
\begin{align}
 a_{\mathcal{M} - 1} = \frac{\bra{j,-j}[C]_{K_{j,\ell,[v]}}\ket{j,-j+\mathcal{M}-1}}{\beta_{\mathcal{M} - 1,\mathcal{M} - 1}}
\end{align}
\noindent and to the following recurrence relation:
\begin{align}
    a_{i} = \frac{1}{\beta_{i,i}}
    \Big[\bra{j,-j}[C]_{K_{j,\ell,[v]}}\ket{j,-j+i} - \sum_{i' = i+1}^{\mathcal{M}-1} a_{i'} \beta_{i',i}\Big].
\end{align}
\noindent Given this relation, constructing $P$ is straightforward. The same then goes for the eigenvalues of $C$ and the entanglement entropy.

\section{Concluding remarks}

We have considered systems of free fermions on Hamming graphs with $d + 1 $ parameters. We have recalled the connection between $\mathfrak{su}(2)$ and Krawtchouk polynomials. We have provided analytical expressions for the entanglement entropy of subsystems corresponding to Hamming graphs of a lower dimension. Similar results were also obtained for subsystems corresponding to a small number of neighborhoods. This was made possible by the decomposition of Hamming graphs in chains, i.e. the identification of the irreducible modules of the Terwilliger algebra of the scheme. Finally, we have shown how to construct block-tridiagonal operators commuting with the chopped correlation matrix that proved of great assistance in the determination of the entanglement entropy for subsystems consisting in a large number of neighborhoods. 

The approach we used rested on the fact that the Hamming graphs are distance-regular or equivalently, related to a P- and Q- polynomial association scheme. Exploring the entanglement entropy of free fermions on graphs of other schemes would also be warranted. Another well known family of graphs are those of the Johnson scheme whose eigenvalue matrices are given in terms of the dual Hahn polynomials. We plan on examining the entanglement entropy of free fermions on these graphs and expect to report the results soon. In the future, it may also be interesting to consider the dual polar and Grassmann graphs \cite{brouwer2012distance}, which are also distance-regular and related to $q$-polynomials of the Askey-scheme \cite{thiniii}.

\section*{Acknowledgements}

We thank Krystal Guo for discussions. PAB holds a scholarship from the Natural Sciences and Engineering Research Council of Canada (NSERC). The research of LV is supported in part by a Discovery Grant from NSERC.

\newpage

\end{document}